\documentclass[10pt]{article}

\usepackage{amsmath, amsthm, latexsym, amssymb, bm, amsfonts,
mathbbol}

\renewcommand{\v}[1]{\mathbf{#1}}
\newcommand{\bl}[0]{\bullet}

\newcommand{\smallf}[2]{ {\textstyle \frac{#1}{#2}} }

\newcommand{\ZZ}[0]{\mathbb{Z}}
\renewcommand{\Re}[0]{\mathfrak{Re}\!~}
\renewcommand{\Im}[0]{\mathfrak{Im}\!~}
\newcommand{\li}[0]{\text{lin}}
\newcommand{\ro}[0]{\text{rot}}
\newcommand{\PET}[0]{Poincar\'e Equivalence Theorem}
\newlength{\minus}
\newcommand{\ms}[0]{\hspace{\minus}}

\newenvironment{mat}[1]{\begin{array}{@{}*{#1}{r@{}l}@{}}}{\end{array}}
\DeclareMathOperator{\tr}{tr}
\theoremstyle{plain}
\newtheorem{theorem}[subsection]{Theorem}

\newtheorem{proposition}[subsection]{Proposition}

\begin{document}

\title{Characteristic Parameters in Integrated Photoelasticity: An
Application of Poincar\'e's Equivalence Theorem}
\author{Hanno Hammer \thanks{email: H.Hammer@umist.ac.uk} \\
Department of Mathematics, UMIST, \\ PO Box 88,
Manchester M60 1QD, UK}

\date{\today}

\maketitle

\begin{abstract}
The Poincar\'e Equivalence Theorem states that any optical element
which contains no absorbing components can be replaced by an
equivalent optical model which consists of one linear retarder and one
rotator only, both of which are uniquely determined. This has many
useful applications in the field of Optics of Polarized Light. In
particular, it arises naturally in attempts to reconstruct spatially
varying refractive tensors or dielectric tensors from measurements of
the change of state of polarization of light beams passing through the
medium, a field which is known as Tensor Tomography. A special case is
Photoelasticity, where the internal stress of a transparent material
may be reconstructed from knowledge of the local optical tensors by
using the stress-optical laws. -- We present a rigorous  approach to
the Poincar\'e Equivalence Theorem by explicitly proving a matrix
decomposition theorem, from which the Poincar\'e Equivalence Theorem
follows as a corollary. To make the paper self-contained we supplement
a brief account of the Jones matrix formalism, at least as far as
linear retarders and rotators are concerned. We point out the
connection between the parameters of the Poincar\'e-equivalent model
to previously introduced notions of the {\it Characteristic
Parameters} of an optical model in the engineering
literature. Finally, we briefly illustrate how characteristic
parameters and Poincar\'e-equivalent models naturally arise in
Photoelasticity.
\vspace{1em}


\end{abstract}


%
%
\section{Introduction}

The propagation of an electromagnetic wave through a material medium
is the result of the interaction between the fundamental fields
$\v{E}$ and $\v{B}$ with a macroscopic number of microscopic sources
constituting the bulk matter. In principle, these sources must be
incorporated into the dynamics of the total system by appropriate
interaction terms. However, if the frequency bandwidth of the light
under consideration is sufficiently far away from the resonance
frequencies of the macroscopic medium, any photon-atom encounter is
only transient; in this case we may refer to the medium as a passive
optical element. Its macroscopic effect on impinging radiation may be
summarized by introducing a (space- and time-dependent) refractive
index.

There are essentially two classes of such optical elements: Firstly,
polarizers, which absorb, or at least attenuate, one of two given
orthogonal  polarization forms of the light passing through the
medium; and secondly, retarders, which introduce a phase retardation
between the components of these polarization forms, but otherwise
preserve the total intensity of the light. In principle, a polarized
light beam can be analyzed with respect to any two orthogonal
polarization forms. However, in practice, two cases are of particular
importance: The linear retarder, which introduces a phase lag between
the components of linearly analyzed light, and the rotator, which does
the same with respect to the circularly polarized components of a
polarized light beam.

The description of the passage of light through optical elements
depends on whether the light is polarized or not. For unpolarized or
partially polarized light, a description in terms of the Stokes
parameters \cite{StokesPara} (for a definition, see
e.g. \cite{BornWolf}) is appropriate; if these are arranged into
four-dimensional real vectors such that optical elements act on it by
means of real $(4,4)$ matrices we speak of the {\it Mueller formalism}
\cite{Mueller1943,Mueller1948}. If, on the other hand, the light beam
is completely polarized, the polarization state of the transverse
degrees of freedom of the electric field can be conveniently arranged
into a two-dimensional complex vector, while optical elements now act
on it by means of complex matrices. This formalism is called the {\it
Jones calculus} \cite{Jones1941a, Jones1941b, Jones1941c, Jones1942a,
Jones1947a, Jones1947b, Jones1948a, Jones1956a, Jones1956b}. The
above-mentioned concepts prove to be useful not only within a
classical context, but appropriate generalizations are used in Quantum
Optics as well: The unitary representation of beam splitters in the
Jones matrix formalism acting on quantum-mechanical mode operators was
discussed in \cite{CamposEA-PRA1989}. A broad introduction to optical
elements in linear optical networks within a quantum-optical context
was recently given in \cite{Leonhardt-ROP66}.

Within the framework of the Jones calculus, polarizers are described
by Hermitean matrices, while retarders are represented by unitary
ones. The latter reflects the fact that retarders preserve the total
intensity of the light. The fact that unitary matrices have the group
property then implies that an arbitrary sequence of retarders is again
described by a unitary matrix. Such a sequence could be a pile of
birefringent plates, each with constant rotation of principal axes and
constant phase retardation. But we can also consider the limit of
infinitely many, infinitely thin, birefringent plates: This  situation
describes a medium which exhibits a spatially varying dielectric
tensor $\epsilon_{ij}$, or more generally, refractive tensor
$n_{ij}$. In this case, the phase velocity depends not only on the
location within the medium, but also on the polarization state of the
(locally-plane) wave at this point. This corresponds to the most
general case of an inhomogeneous anisotropic medium.

Such a scenario arises naturally in the field of {\it Photoelasticity}
\cite{Frocht1, Frocht2, CokerFilon, TheocarisEA, Ditchburn, Aben1979,
AbenGuillemet, AbenEA1989a}: Experience shows that certain materials,
such as glasses and polymers, are optically isotropic and homogeneous
when unloaded, but exhibit local anisotropy when strained by an
external load. The relation between the resulting dielectric tensor
$\epsilon_{ij}$ and the stress tensor $\sigma_{ij}$ in the interior of
the medium is called the {\it stress-optical law}; its basic form has
been discovered long ago by Maxwell \cite{MaxwellStrOpt}. {\it
Integrated Photoelasticity} is concerned with the reconstruction of
optical tensors, and via the stress-optical law, of stress tensors,
from data sets which are acquired by sending polarized light through
the loaded specimen at many different angles, thereby  measuring the
change in the state of polarization. For this method to work, the
material must be sufficiently transparent to ignore absorption within
the medium. Since tensorial quantities are reconstructed, these
methods belong to the field of {\it Tensor Tomography}.

In the examples above, the medium is given by a pile of retarders
and/or rotators, possibly in the infinitesimal limit. The previous
discussion implies that, within the Jones formalism, such a medium may
be described by a unitary (unimodular) matrix $U$. It turns out that
such a matrix always determines -- and, in turn, is fully determined
by -- a set of {\it characteristic directions} \cite{Aben1966a} and a
{\it characteristic phase retardation} between them. These {\it
characteristic parameters} \cite{Aben1966a} have been introduced
within the context of Photoelasticity from an engineering point of
view, and are not necessarily standard in the mathematical literature
[a rigorous mathematical definition will be given in section
\ref{CharacteristicParameters}]. If the matrix $U$ is interpreted as
belonging to an optical element of the kind as discussed above, the
characteristic parameters can be given an operational meaning: The
{\it primary characteristic directions} determine those planes of
linear polarization at the entry into the medium for which the state
of polarization at the emergence from the medium is again linear. The
{\it secondary characteristic directions} determine the planes of
linear polarization of the emerging light, if the incident light was
linearly polarized in the primary characteristic directions; in
general, they differ from the primary ones. It turns out [sections
\ref{CharacteristicParameters}, \ref{Relation}] that there are always two
orthogonal primary and two orthogonal secondary characteristic
directions. Light which is linearly polarized along the two primary
directions travels with different phase velocities, though, so that
both waves emerge with a phase difference -- the characteristic phase
retardation.

It is an important task to reconstruct the optical properties of the
medium, i.e. the matrix $U$, from measurements of the characteristic
parameters. To this end, relations between the parameters of a normal
form of unitary matrices and the characteristic parameters may be
derived \cite{Aben1979, AbenGuillemet, Aben1966a}. However, these
relations only determine the squares of sines or cosines of angles and
therefore the actual computation of these quantities is rather
involved. It is at this point where the {\it Poincar\'e Equivalence
Theorem} \cite{Poincare1892} comes into play: This theorem states that
any optical medium which is described by a unitary unimodular matrix
can be replaced by an optically equivalent model which consists of one
retarder followed by one rotator, or the other way round; in each
case, both elements are uniquely determined. Then it can be shown
easily that the characteristic parameters of the optical medium
coincide with the optical  parameters of the equivalent model, i.e.
with  the principal directions and phase retardation for the
equivalent retarder, and the rotation angle for the equivalent
rotator; but from the latter, the matrix $U$ may readily be determined
completely. This reconstruction becomes particularly important in the
case of tensor tomography of photoelastic media, where the overall
effect of the specimen on the passage of light represents the
accumulated effect of many (infinitesimal) retarders and rotators.

These examples highlight the importance of the \PET\ in the field of
Optics of Polarized Light. Although references to the equivalence
theorem, or applications thereof, are made frequently in the
literature \cite{TheocarisEA, Aben1979, AbenGuillemet, AbenEA1989a,
TomlinsonPatterson2002a, TomlinsonPatterson1998a}, its precise content
is apparently not widely published; a rigorous proof of the theorem,
or some equivalent statement thereof, seems even less readily
available. This observation constitutes the starting point of our
paper: We rigorously prove a matrix decomposition theorem for unitary
unimodular $(2,2)$ matrices [section \ref{deco}], from which the \PET\
follows as a corollary [section \ref{PoiEqu}]. To put this result into
a context, and to make the paper reasonably self-contained, we provide
a brief overview of the description of polarized light, and optical
elements acting on it, in terms of the Jones calculus; this is done in
section
\ref{JonesMatrixFormalism}.  The second objective of this paper is to
clarify the relation between the (operationally defined)
characteristic parameters of a transparent model according to Aben
\cite{Aben1979, AbenGuillemet, Aben1966a}, and the optical parameters
of the Poincar\'e-equivalent model. To this end we recapitulate in
section
\ref{CharacteristicParameters} the notions of characteristic
directions and characteristic phase retardation for non-absorbing
optical elements. We also prove the existence of characteristic
directions for any unitary unimodular $(2,2)$ matrix in a standard
parametrisation. The  relation between characteristic parameters and
the parameters of the Poincar\'e-equivalent retarder and rotator then
are analyzed in section \ref{Relation}. Finally, in section
\ref{Photoelasticity} we outline the basic ideas of three-dimensional
photoelasticity in order to illustrate the context in which a specimen
with continuous variation of optical tensors presents itself, and how
the Poincar\'e-equivalent model may be important to the objective of
tensor tomography. In section \ref{Summary} we summarize our results.

\section{Polarized Light in the Jones Matrix Formalism}
\label{JonesMatrixFormalism}

In this paper we only consider {\it completely polarized} classical
light. Strictly speaking, this is an idealization, which can be
approximately realized only if 1) the light is strictly monochromatic,
and 2) the {\it complex degree of coherence} \cite{BornWolf} of two
orthogonal components of the electric field always has unit
modulus. Physically, this means that the two orthogonal components
have a well-defined and constant phase relation for all times.

The transverse electric field of a general eliptically polarized light
beam propagating in the $3$-direction takes the form
\begin{equation}
\label{Jones1}
 \v{E}(\v{x},t) = a_1 \, \cos\left(kz-\omega t + \delta_1\right) \,
 \v{e}_1 + a_2 \, \cos\left(kz-\omega t + \delta_2\right) \, \v{e}_2
 \quad,
\end{equation}
where $a_1, a_2$ are real amplitudes, $u=\omega/k$ is the phase
velocity in the (isotropic and homogeneous) medium, $\delta_1$ and
$\delta_2$ are constant phases, and $\v{e}_1$ and $\v{e}_2$ are real
unit vectors in the direction of the $x$- and $y$-axes. Since the
action of optical elements on light beams is accomplished by linear
transformations, it is admissible to assume that the beam has unit
relative  intensity \cite{Ditchburn} $a_1^2+a_2^2=1$. We now choose a
fixed  location within the beam, say, $z=0$, and express
(\ref{Jones1}) as the real part of the complex vector
\begin{equation}
\label{Jones2}
 \v{E} = e^{-i\omega t} \left\{ a_1 \, e^{i\delta_1} \, \v{e}_1 + a_2
 \, e^{i\delta_2} \, \v{e}_2 \right\} \quad.
\end{equation}
The quantity in curly brackets defines the {\it polarization form} of
the light; the associated complex $2$-vector
\begin{equation}
\label{Jones3}
v_l =\left( \begin{mat}{2} v_1 \\ v_2 \end{mat} \right) \equiv \left(
\begin{mat}{1} a_1 \, e^{i\delta_1} \\ a_2 \, e^{i\delta_2} \end{mat}
\right) \quad,
\end{equation}
is called the {\it Jones vector} of the light beam with respect to
linear polarization along the directions $\v{e}_1$ and $\v{e}_2$. We
use a subscript $l$ to denote the fact that the Jones vector
(\ref{Jones3}) denotes linearly polarized components.

As can be seen, the polarization form and its associated Jones vector
are defined only up to a global phase, i.e. Jones vectors $v$ and
$e^{i\Phi} v$ are equivalent, since the global phase $\Phi$ can always
be reabsorbed into the total phase $e^{-i\omega t}$ governing the time
dynamics.

Eq. (\ref{Jones2}) represents the decomposition of the (complex form
of the) electric field vector into two linearly polarized components
whose amplitudes are $a_1, a_2$, and whose relative phase is
$\delta\equiv \delta_2-\delta_1$. This decomposition is convenient to
derive the Jones matrix of a linear retarder. To study the rotator it
is necessary to analyze the light with respect to complex basis
vectors $\v{e}_+$ and $\v{e}_-$ which represent right-handed and
left-handed circular polarization, respectively, which are defined as
\begin{subequations}
\label{rot1}
\begin{align}
\v{e}_{\pm} & = \frac{1}{\sqrt{2}} \left( \v{e}_1 \pm i \v{e}_2 \right)
\quad, \label{rot1a} \\
(\v{e}_+ \;,\; \v{e}_-) & = (\v{e}_1 \;,\; \v{e}_2) \, \v{M} \quad,
\quad
\v{M} = \frac{1}{\sqrt{2}} \left( \begin{array}{@{}*{2}{r@{}l}@{}} &1&&1 \\
&i&-&i \end{array} \right)  \label{rot1b} \quad.
\end{align}
\end{subequations}
[We recall that right-handed = left-circular, and vice versa]. The
electric field vector (\ref{Jones2}) can be expanded in terms of the
basis $(\v{e}_+,
\v{e}_-)$,
\begin{equation}
\label{Jones4}
 \v{E} \sim (\v{e}_1 \;,\; \v{e}_2) \, \left( \begin{mat}{2} v_1 \\
 v_2 \end{mat} \right) = (\v{e}_+ \;,\; \v{e}_-) \, \left(
 \begin{mat}{2} v_+ \\ v_- \end{mat} \right) \quad,
\end{equation}
such that the components $v_+$ and $v_-$ make up the Jones vector
$v_c$ of the light in the basis (\ref{rot1}),
\begin{equation}
\label{Jones5}
 v_c = \left( \begin{mat}{1} v_+ \\ v_- \end{mat} \right) =
 \v{M}^{\dag} \, v_l = \frac{1}{\sqrt{2}} \left( \begin{mat}{2} &1&-&i
 \\ &1&&i \end{mat} \right) \left( \begin{mat}{2} v_1 \\ v_2 \end{mat}
\right) \quad.
\end{equation}
We use a subscript $c$ to point out that the Jones vector
(\ref{Jones5}) denotes circularly polarized components.

Evidently, the matrix $\v{M}$ mapping the basis vectors $(\v{e}_1,
\v{e}_2)$ for linear polarization to those $(\v{e}_+, \v{e}_-)$ for
circular polarization is unitary. Indeed, this is true for any
transformation between basis vectors of different polarization forms. 

We are now in a position to derive the Jones matrices for linear
retarders and rotators.

\subsection{The linear retarder in the Jones calculus}

A {\it linear retarder} is an optical element made of a material which
exhibits homogeneous optical anisotropy along a given direction of
light transmission \cite{BornWolf, TheocarisEA,
Ditchburn}. Perpendicular to the direction of passage of light, the
retarder has two distinct orthogonal directions, called the {\it fast}
and the {\it slow} axis, respectively. The phase velocity of a plane
wave which is linearly polarized along the fast axis is greater than
for waves polarized along the slow axis. This means that, at the point
of emergence from the retarder, the component along the slow axis
(''slow component'') has acquired a phase lag $\delta$ with respect to
the component along the fast axis (''fast component''). Let us assume
that the fast/slow axis is oriented along the $x/y$-axis. If the Jones
vector of light prior to entry into the retarder is
\begin{equation}
\label{ret1}
v = \left( \begin{mat}{1} c_1 \\ c_2 \end{mat} \right) \quad,
\end{equation}
then after emergence from the retarder it has changed to
\begin{equation}
\label{ret2}
v' = e^{i\Phi_g} \left( \begin{mat}{1} c_1 \\ e^{i\delta} \, c_2 
\end{mat}
\right) \quad.
\end{equation}
The exponential $e^{i\delta}$ indeed represents a phase {\it lag} of
the local harmonic oscillator at $z=0$ vibrating in the $y$-direction,
because of our convention to count the phase of the time dynamics as
$e^{-i\omega t}$. Furthermore, $\Phi_g$ represents a global phase
which is picked up by any wave passing through the retarder,
irrespective of its polarization form. Since global phases are
immaterial in our present discussion, it is common to adjust them for
maximum convenience, or otherwise discard them altogether, which is
what we shall do throughout. Physically, a global change of phase
could be accomplished by passage of the light through an {\it
isotropic} medium whose optical path length corresponds to the desired
phase difference.

Since the global phase in (\ref{ret2}) can be chosen freely, we may
extract and discard a global phase $\Phi_g = \smallf{\delta}{2}$, in
which case the transformed Jones vector reads
\begin{equation}
\label{ret2a}
v' = \left( \begin{mat}{1} e^{-i\delta/2} \, c_1 \\ e^{i\delta/2} \,
c_2 \end{mat}
\right) \quad.
\end{equation}
It is now clear that the matrix which accomplishes the transformation
$v \rightarrow v'$ must take the form
\begin{equation}
\label{ret3}
 \v{J}_{\li}(0,\delta) = \left( \begin{mat}{2} & e^{-i\delta/2} & & 0
 \\ & 0 & & e^{i\delta/2} \end{mat} \right) \quad.
\end{equation}
Our notation $\v{J}_{\li}(0,\delta)$ reflects the fact that the phase
difference between the components is $\delta$, while the angle between
the fast axis and the $x$-direction is $\theta=0$.

It is easy to derive the Jones matrix $\v{J}_{\li}(\theta, \delta)$ of
a linear retarder whose fast axis makes a nonvanishing angle with the
$x$-axis. To this end we introduce the {\it rotation matrix}
\begin{equation}
\label{linPol4}
 \v{R}(\theta) = \left( \begin{array}{@{}*{2}{r@{}c}@{}} & \cos\theta & &
 \sin\theta \\ -& \sin\theta & & \cos\theta \end{array} \right) \quad.
\end{equation}
This is an $SO(2)$-matrix representing a passive rotation by $\theta$
about the $z$-axis, that is to say, the basis vectors and vector
components transform according to
\begin{subequations}
\label{rota1}
\begin{align}
 \v{E}  \sim (\v{e}_1 \;,\; \v{e}_2) \left( \begin{mat}{1} v_1 \\ v_2
 \end{mat} \right) = (\v{e}'_1 \;,\; \v{e}'_2) \left( \begin{mat}{1}
 v'_1 \\ v'_2 \end{mat} \right) \quad & , \quad (\v{e}'_1 \;,\;
 \v{e}'_2) = (\v{e}_1 \;,\; \v{e}_2) \, \v{R}(\theta)^T \quad,
\label{rota1a} \\
 \left( \begin{mat}{1} v'_1 \\ v'_2 \end{mat} \right)  = \v{R}(\theta)
 \, \left( \begin{mat}{1} v_1 \\ v_2 \end{mat} \right) \quad & , \quad
 \v{R}(\theta)^T = \v{R}(\theta)^{-1} = \v{R}(-\theta)
 \quad. \label{rota1b}
\end{align}
\end{subequations}
The group property implies that a succession of two rotations is again
a rotation; furthermore, two rotations around the same axis
commute. Thus we have
\begin{align}
 \v{R}(\theta_1+\theta_2) & = \v{R}(\theta_1) \, \v{R}(\theta_2) =
 \v{R}(\theta_2) \, \v{R}(\theta_1) \quad. \label{linPol5a} 
\end{align}
Using the rotation matrices it is easy to derive the Jones matrix
$\v{J}_{\li}(\theta, \delta)$ for a linear retarder with nonvanishing
angle $\theta$: We only need to keep in mind that in the reference
frame of the fast ($\v{e}'_1$) and slow ($\v{e}'_2$) axes, the
retarder must act with a Jones matrix (\ref{ret3}) on the components
$v'_1$ and $v'_2$. Thus,
\begin{equation}
\label{ret4}
\left.
\begin{aligned}
 \v{E} & \sim (\v{e}'_1 \;,\; \v{e}'_2) \left( \begin{mat}{1} v'_1 \\
 v'_2 \end{mat} \right) \rightarrow (\v{e}'_1 \;,\; \v{e}'_2) \,
 \v{J}_{\li}(0,\delta) \, \left( \begin{mat}{1} v'_1 \\ v'_2 \end{mat}
 \right)  \\ & = (\v{e}_1 \;,\; \v{e}_2) \, \v{R}(\theta)^T \,
 \v{J}_{\li}(0,\delta) \, \v{R}(\theta) \, \left( \begin{mat}{1} v_1
 \\ v_2 \end{mat} \right) = (\v{e}_1 \;,\; \v{e}_2) \,
 \v{J}_{\li}(\theta, \delta) \, \left( \begin{mat}{1} v_1 \\ v_2
 \end{mat} \right) \quad,
\end{aligned} \qquad
\right\}
\end{equation}
where we have used eqs. (\ref{rota1}). It follows that
\begin{equation}
\label{ret5}
 \v{J}_{\li}(\theta, \delta) = \v{R}(-\theta) \,
 \v{J}_{\li}(0,\delta) \, \v{R}(\theta) \quad,
\end{equation}
or explicitly,
\begin{equation}
\label{ret5a}
 \v{J}_{\li}(\theta, \delta) = \left( \begin{array}
 {@{}*{2}{r@{}r}@{}} 
 \cos\frac{\delta}{2} & - i \, \sin\frac{\delta}{2} \, \cos2\theta &
 & -i \, \sin\frac{\delta}{2} \, \sin2\theta \\[3pt]
 & -i \, \sin\frac{\delta}{2} \, \sin2\theta & \ms\ms
 \cos\frac{\delta}{2} & + \sin\frac{\delta}{2} \, \cos2\theta 
 \end{array} \right) 
 \quad.
\end{equation}

Equation (\ref{ret5}) is indeed valid for Jones matrices $\v{J}$ of
arbitrary optical devices whose preferred axes are related by an
orthogonal transformation, since all we have used in the derivation of
(\ref{ret5}) was the law of passive coordinate transformations onto a
coordinate system whose axes coincided with the preferred axes of the
optical device.

We note that the Jones matrices of the linear retarder, (\ref{ret3})
and (\ref{ret5}), are unitary unimodular matrices, since we have
chosen the global phase $\Phi_g$ in (\ref{ret3}) in such a way that
$\det\v{J}_{\li}(0,\delta)=1$.

\subsection{The rotator in the Jones calculus}

The rotator is an optical element which introduces a phase lag
$\delta$ of the left-handed relative to the right-handed component of
the complex electric field vector (\ref{Jones2}). If the electric
field is expanded in terms of the basis (\ref{rot1}), the left-handed
/ right-handed components are just the elements $v_-$ and $v_+$ in the
Jones vector $v_c$ given in eq. (\ref{Jones5}). Thus, after passage
through the rotator we have
\begin{equation}
\label{rota11}
 v_c = \left( \begin{mat}{1} v_+ \\ v_- \end{mat} \right)
 \longrightarrow v'_c = \left( \begin{mat}{1} e^{-i\delta/2} \, v_+ \\
 e^{i\delta/2} \, v_- \end{mat} \right) \quad.
\end{equation}
Evidently, the linear map which accomplishes the transformation
(\ref{rota11}) is precisely the matrix $\v{J}_{\li}(0,\delta)$ of the
linear retarder given in (\ref{ret3}); thus, in the basis
(\ref{rot1}) we simply have
\begin{equation}
\label{rota2}
 v'_c = \v{J}_{\li}(0,\delta) \; v_c \quad.
\end{equation}
Using eq. (\ref{Jones5}) which effects the transformation between the
components for linear and circular polarization, we can easily
transform eq. (\ref{rota2}) back into the Cartesian frame $(\v{e}_1,
\v{e}_2)$, in which the Jones vector is given by (\ref{Jones3}),
\begin{equation}
\label{rota3}
 v'_l = \v{M} \, \v{J}_{\li}(0,\delta) \, \v{M}^{\dag} \; v_l =
 \v{J}_{\ro}(-\smallf{\delta}{2}) \; v_l \quad.
\end{equation}
This determines the Jones matrix $\v{J}_{\ro}(-\smallf{\delta}{2})$ of
the rotator in the Cartesian basis,
\begin{equation}
\label{rota4}
 \v{J}_{\ro}(-\smallf{\delta}{2}) = \left(
 \begin{array}{@{}*{2}{r@{}l}@{}} & \cos\frac{\delta}{2} & - &
 \sin\frac{\delta}{2} \\[3pt] & \sin\frac{\delta}{2} & &
 \cos\frac{\delta}{2} \end{array} \right) = \v{R}(-\smallf{\delta}{2})
 \quad,
\end{equation}
where the rotation matrix $\v{R}$ was defined in (\ref{linPol4}).

The reason for our seemingly strange notation convention becomes clear
if we recall that a positive phase lag of the left-handed component
with respect to the right-handed component should rotate the plane of
polarization of a linearly polarized light beam in a counter-clockwise
sense; this is a positive angle $+\smallf{delta}{2}$ in the
$xy$-plane, if we look {\bf into} the oncoming wave, i.e. from
positive towards negative $z$-values. It is easy to check this:
\begin{equation}
\label{rotJon7}
 \v{J}_{\ro}(-\smallf{\delta}{2}) \, \left[ \begin{array}{@{}c@{}}
 \cos\theta \\ \sin\theta
 \end{array} \right] = \left[ \begin{array}{@{}c@{}} \cos \left(
 \theta + \frac{\delta}{2} \right) \\ \sin \left( \theta +
 \frac{\delta}{2} \right) \end{array} \right] \quad.
\end{equation}
Here, $[\cos\theta, \sin\theta]^T$ is the Jones vector of a
normalized, linearly polarized beam, and $\theta$ is the initial angle
of inclination of the plane of polarization. After passage through the
rotator, the plane of polarization has rotated by $+\delta/2$.

This finishes our discussion of linear retarders and rotators in the
Jones calculus.

\section{Characteristic parameters of $SU(2)$ matrices}
 \label{CharacteristicParameters}

As mentioned in the Introduction, the terms 'Characteristic
Directions', 'Characteristic Phase Difference' and 'Characteristic
Parameters' were first introduced by Aben within the framework of
Photoelasticity \cite{Aben1966a}. We now wish to explain their
meaning.

Consider a non-absorbing optical medium whose Jones matrix $U$ is
unitary. If $U$ happens to have a determinant different from $1$, we
can always extract and discard a global phase, so that $U$ may be
assumed to be unimodular, $\det U = 1$, and hence an element of
$SU(2)$. A parametrisation of $SU(2)$ matrices is given by
\cite{Cornwell1}
\begin{equation}
\label{char1}
U = \left( \begin{array}{@{}*{2}{r@{}l}@{}} &\cos\theta \, e^{i\phi}
 &-& \sin\theta \, e^{i\chi} \\ &\sin\theta \, e^{-i\chi} & &
 \cos\theta \, e^{-i\phi} \end{array} \right) \quad, \quad \theta \in
 [0,\smallf{\pi}{2}] \quad, \quad \phi, \chi \in [0,2\pi) \quad.
\end{equation}
This parametrisation is unique in the interior of the interval
$[0,\smallf{\pi}{2}]$, i.e. for $0 < \theta < \smallf{\pi}{2}$. If $\theta$
takes one of the boundary values, parts of the parameter space must
undergo identification in order to make the map between parameters and
matrices unique: For $\theta=0$ we must identify points $(\theta,
\phi, \chi) = (0,0,\chi)$, $\chi \in [0,2\pi)$. For $\theta=\smallf{\pi}{2}$,
the points $(\theta, \phi, \chi) = (\smallf{\pi}{2}, \phi, 0)$, $\phi \in
[0,2\pi)$, must be identified. This agreement makes the mapping
between parameter space and matrices bijective.

Using this parametrisation we prove the
\begin{theorem}[Characteristic directions]
\label{CharacteristicDirections} 

For every unimodular unitary matrix $U$ of the form
(\ref{char1}) there exist two {\em real} vectors
\begin{equation}
\label{char2}
 w_m = \left( \begin{array}{@{}c@{}} \cos\gamma_m \\
 \sin\gamma_m \end{array} \right) \quad, \quad m=1,2 \quad,
\end{equation}
such that the vectors $U w_m$ are {\em real} vectors times an overall
phase,
\begin{equation}
\label{char3}
U w_m = e^{i\Phi_m} \left(\begin{array}{@{}c@{}} \cos\gamma'_m \\
 \sin\gamma'_m \end{array} \right) = e^{i\Phi_m} \, w'_m  \quad, \quad
 m=1,2 \quad.
\end{equation}
The angles $\gamma_m$ are solutions of the equation
\begin{equation}
\label{char4}
 \tan 2\gamma_m = - \frac{\sin(\phi+\chi) \sin2\theta}{\cos^2\theta
 \sin2\phi - \sin^2\theta \sin 2\chi} \quad,
\end{equation}
such that
\begin{equation}
\label{char5}
 \gamma_1 \in [-\smallf{\pi}{4}, \smallf{\pi}{4}] \quad \text{and}
 \quad \gamma_2 = \gamma_1 + \smallf{\pi}{2} \quad.
\end{equation}
The vectors $w_m, m=1,2$ as given in (\ref{char2}) span the {\em
primary characteristic directions} of the matrix $U$.

The angles $\gamma'_m$ are solutions of the equation
\begin{equation}
\label{char6}
 \tan 2\gamma' =  \frac{\sin(\phi-\chi) \sin2\theta}{\cos^2\theta
 \sin2\phi + \sin^2\theta \sin 2\chi} \quad,
\end{equation}
such that 
\begin{equation}
\label{char7}
 \gamma'_1 \in [-\smallf{\pi}{4}, \smallf{\pi}{4}] \quad \text{and}
 \quad \gamma'_2 = \gamma'_1 + \smallf{\pi}{2} \quad.
\end{equation}
The vectors $w'_m, m=1,2$ as given in (\ref{char3}) span the {\em
secondary characteristic directions} of the matrix $U$.

The angles $\gamma_m$ and $\gamma'_m$ are well-defined if and only if
\begin{subequations}
\label{spec}
\begin{align}
 (\theta,\phi) & \not\in \Big\{\; (0,0) ,\; (0, \smallf{\pi}{2}) ,\; (0, \pi)
,\; (0, \smallf{3\pi}{2})\; \Big\} \quad, \label{spec1} \\ 
 (\theta, \chi) & \not\in \Big\{ \; ( \smallf{\pi}{2},0) ,\; (
 \smallf{\pi}{2} , \smallf{\pi}{2} ) ,\; ( \smallf{\pi}{2}, \pi )
 ,\; ( \smallf{\pi}{2} ,\; \smallf{3\pi}{2} ) \; \Big\}
 \quad. \label{spec2}
\end{align}
\end{subequations}
If one of the parameter values (\ref{spec}) is taken, the right-hand
side of eqs. (\ref{char4}, \ref{char6}) becomes the ill-defined
expression $\smallf{0}{0}$, and no angles $\gamma_m$, $\gamma'_m$ can
be determined.
\end{theorem}
{\it Proof:}

If the statement (\ref{char3}) holds, then the quotient $w'_{m1} /
w'_{m2}$, or equivalently, the product $w'_{m1} (w'_{m2})^*$, must be
real, for $m=1,2$. This requires that
\begin{equation}
\label{pro1}
\mathfrak{Im} \; w'_{m1} \, \left( w'_{m2} \right)^* = 0 \quad.
\end{equation}
We now express $(w'_m)_{1,2}$ as $(U w_{m})_{1,2}$, respectively, and
impose the condition (\ref{pro1}). Under the conditions (\ref{spec})
this leads to eq. (\ref{char4}), which determines an angle $2\gamma_1$
lying between $-\smallf{\pi}{2}$ and $+\smallf{\pi}{2}$. Then, the
angle $2\gamma_1+\pi$ also satisfies (\ref{char4}). It follows that
$\gamma_1$ lies between $-\smallf{\pi}{4}$ and $+\smallf{\pi}{4}$, and
the second solution is $\gamma_2 = \gamma_1+\smallf{\pi}{2}$. Thus,
there are two primary characteristic directions, and they are
orthogonal.

To determine the angles $\gamma'_m$, we invert (\ref{char3}),
\begin{equation}
\label{pro2}
 U^{\dag} w'_m = e^{-i\Phi_m} \, w_m  \quad.
\end{equation}
This has the form of eq. (\ref{char3}), and we can therefore use the
same line of arguments as before, if we put the matrix $U^{\dag}$ into
the same form as (\ref{char1}),
\begin{equation}
\label{pro3}
U \rightarrow U^{\dag} \quad \Rightarrow \quad \theta \rightarrow
-\theta, \quad \phi \rightarrow -\phi , \quad \chi \rightarrow \chi
\quad.
\end{equation}
If the replacements (\ref{pro3}) are used in (\ref{char4}) we obtain
(\ref{char6}), again under the conditions (\ref{spec}). --- It is easy
to check that, whenever one of the values in (\ref{spec}) is taken,
the right-hand sides in both (\ref{char4}) and (\ref{char6}) are
ill-defined. --- This finishes the proof. \hfill $\blacksquare$
\vspace{0.5em}

The physical meaning of this is as follows: The existence of two
solutions for (\ref{char3}) implies that there exist two perpendicular
directions for the plane of linearly polarized light, given by the
solutions (\ref{char2}), such that if light polarized in this
direction  enters the device it will emerge linearly polarized again;
this is a consequence of the fact that the components of the
transformed Jones vector have the same overall phase. This statement
is true only for the light as it emerges {\it behind} the apparatus;
in the interior of the device, the light is elliptically polarized in
general. Furthermore, the phase velocity of a linearly polarized wave
entering the device at the angle $\gamma_1$ differs from the one
entering at $\gamma_2$, so that both waves will emerge with a phase
difference $\Delta$, see eq. (\ref{Phdiff0}) below.

We now turn to the determination of the phases $\Phi_m$ in
(\ref{char3}). We find
\begin{proposition}[Determination of phases]
\label{DeterminationPhases}
The phases $\Phi_m$, $m=1,2$ satisfy the equations
\begin{subequations}
\label{char3b}
\begin{align}
 \cos^2\Phi_m & = \cos^2\theta \, \cos^2\phi + \sin^2\theta \,
 \cos^2\chi \quad, \label{char3ba} \\ 
 \sin^2\Phi_m & = \cos^2\theta \, \sin^2\phi + \sin^2\theta \,
 \sin^2\chi \quad, \label{char3bb} \\
 \tan^2\Phi_m & = \frac{\sin^2\phi + \sin^2\chi \,
 \tan^2\theta}{\cos^2\phi + \cos^2\chi \, \tan^2\theta} \quad.
 \label{char3bc}
\end{align}
\end{subequations}
and
\begin{equation}
\label{char3c}
 \cos2\Phi_m = \cos^2\theta \, \cos2\phi + \sin^2\theta \, \cos2\chi
 \quad.
\end{equation}
\end{proposition}
{\it Proof:}

We perform the matrix multiplication $U w_m$ with the form
(\ref{char1}) for $U$, but the angles $\gamma_m$ need not be known
explicitly. On the right-hand side of this equation we have
$e^{i\Phi_m} \, w'_m$. We now expand this equation into real and
imaginary parts, obtaining two equations which on their right-hand
sides contain $\cos\Phi_m \, w'$ and $\sin\Phi_m \, w'$,
respectively. If the component equations are squared and added, we
obtain the relations (\ref{char3ba}, \ref{char3bb}).  To arrive at
(\ref{char3bc}), we divide (\ref{char3ba}) by (\ref{char3bb}) and then
divide the whole equation by a common factor $\cos^2\theta$.
Subtracting (\ref{char3bb}) from (\ref{char3ba}) then yields
(\ref{char3c}). \hfill$\blacksquare$\vspace{0.5em}

It is important to point out that the phase changes $\Phi_m$ in
general do {\it not} coincide with the real phases of the light as it
passes through the real device, for in the latter case there is also
an additional  global phase which is accumulated by any light beam
that travels through the apparatus. This global phase has been
discarded in the construction of the Jones matrices of the retarder
and the rotator, in order to make these matrices unimodular. However,
the global phase is the same for light of any polarization form, and
therefore cancels out if we construct the phase {\it difference}
between light beams which enter the device along the primary
characteristic directions. This phase difference does not depend on
unobservable parameters and can therefore be measured. In principle,
it can be evaluated from (\ref{char3bc}): This equation has four
solutions, given by
\begin{equation}
\label{Phdiff1}
 \Phi \in [-\smallf{\pi}{2},\smallf{\pi}{2}] \quad, \quad \Phi+\pi
 \quad, \quad -\Phi \quad, \quad -\Phi+\pi \quad.
\end{equation}
One of the solutions $\Phi_1$ and $\Phi_2$ must be chosen from the set
$\left\{ \Phi, \Phi + \pi \right\}$, while the second one must take
values in $\left\{ -\Phi, -\Phi + \pi \right\}$. In both cases we
arrive at the conclusion that
\begin{equation}
\label{Phdiff0}
  \Delta = 2 \Phi_2  \mod \pi \quad.
\end{equation}
At this point a remark concerning the results (\ref{char7}) and
(\ref{Phdiff0}) are in order:

Eq. (\ref{char6}) determines the directions of polarization of the
emerging light. However, it does not determine whether $U  w_1$ is
equal to $w'_1$ or $w'_2$, and similarly for $U w_2$. This is
reflected in the indeterminacy of the phase difference in
(\ref{Phdiff0}): The modulus of $\Delta$ is determined only up to
$\pi$, while the sign is undetermined. It is at this point where the
decomposition theorem \ref{decoTheo} and the ensuing \PET\ prove very
useful, since they allow to remove this indeterminacy completely,
provided the measurements on the sample are capable of distinguishing
between the fast and the slow axis of the equivalent retarder [see
sections \ref{PoiEqu}, \ref{Relation}].

For every optical device described by a unitary unimodular matrix $U$,
the set of quantities $(\gamma_1, \gamma'_1, \Delta)$, are called the
{\it characteristic parameters} of the model.

\section{Matrix Equivalence and the \PET}
\label{deco}

In theorem \ref{decoTheo} we show that every two-dimensional unitary
unimodular matrix $U$ has a unique decomposition in terms of a minimal
number of 'basic building blocks'. There are two of these building
blocks: The first is given by the rotation matrices $\v{R}(\xi)$ as
defined in (\ref{linPol4}); the second is given by {\it diagonal}
unitary unimodular matrices
\begin{equation}
\label{Poin1b}
 \v{D}(\delta) = \left( \begin{mat}{2} &e^{i\delta}
 &&0\\&0&&e^{-i\delta} \end{mat} \right) \quad, \quad \delta \in
 [0,2\pi) \quad.
\end{equation}
After that, in theorem \ref{PoiEquTheo} we deduce the \PET\ from
theorem \ref{decoTheo}.

{
\renewcommand{\theenumi}{\Roman{enumi}}
\renewcommand{\labelenumi}{\theenumi.)}

\begin{theorem}[Decomposition] \label{decoTheo}
Let $U$ be a unitary unimodular matrix in the parametrisation
(\ref{char1}). Then the following statements are true:
\begin{enumerate}
\item \label{case1} If the angle $\theta$ in (\ref{char1}) lies in the
interior of the interval $[0,\smallf{\pi}{2}]$, $0 < \theta < \smallf{\pi}{2}$,
then there exist angles
\begin{align}
 \alpha , \beta & \in [0,\pi) \quad, \quad \delta \in [0,2\pi)
 \quad, \label{Poin1}
\end{align}
such that $U$ has a decomposition
\begin{subequations}
\label{oeq1}
\begin{align}
 U & = \v{R}(-\smallf{\alpha+\beta}{2}) \, \v{D}(\delta) \,
 \v{R}(\smallf{\alpha-\beta}{2}) \quad \label{oeq0} \\
 & = \left( \begin{array}{@{}*{2}{r@{}l}@{}} & \cos\frac{\alpha+\beta}{2}
 &-&\sin\frac{\alpha+\beta}{2} \\[2pt] &\sin\frac{\alpha+\beta}{2} &
 &\cos\frac{\alpha+\beta}{2} \end{array}  \right)
 \left( \begin{array}{@{}*{2}{r@{}l}@{}} & e^{i\delta} & & 0 \\[2pt] & 0 &
 & e^{-i\delta} \end{array}  \right)
\left(
 \begin{array}{@{}*{2}{r@{}l}@{}} & \cos\frac{\alpha-\beta}{2}
 &&\sin\frac{\alpha-\beta}{2} \\[2pt] -&\sin\frac{\alpha-\beta}{2} &
 &\cos\frac{\alpha-\beta}{2} \end{array}  \right) 
  \label{oeq1a} \\ 
 & = \left( \begin{array}{@{}*{2}{c}@{}} \cos\delta \cos\beta + i
 \sin\delta \cos\alpha & -\cos\delta \sin\beta + i \sin\delta
 \sin\alpha \\[2pt] \cos\delta \sin\beta + i \sin\delta \sin\alpha & \ms
 \cos\delta \cos\beta - i \sin\delta \cos\alpha \end{array}  \right)
 \quad. \label{oeq1b} 
\end{align}
\end{subequations}
The angles (\ref{Poin1}) are uniquely determined, and there is no way
to represent $U$ by less than three matrices of the kind
(\ref{linPol4}) and (\ref{Poin1b}).
\item \label{case2}
For $\theta=0$ we have
\begin{subequations}
\begin{equation}
\label{descri2}
 U = \v{D}(\phi) \quad.
\end{equation}
This is a unique minimal decomposition in terms of \underline{one}
diagonal matrix (\ref{Poin1b}).

A representation in terms of \underline{three} matrices, form
(\ref{oeq0}), may be given for parameter values
\begin{align}
 \alpha & = \beta = 0 \quad, \quad \delta = \phi \quad,
 \label{Poin1bAA}
\end{align}
which is unique provided that
\begin{equation}
\label{Poin1bAB}
\phi \not\in \left\{0, \smallf{\pi}{2}, \pi, \smallf{3\pi}{2} \right\}
\quad.
\end{equation}
\end{subequations}
If $\phi \in \left\{0, \smallf{\pi}{2}, \pi, \smallf{3\pi}{2}
\right\}$, a decomposition in terms of \underline{three} matrices is
not unique, since one of the angles $\alpha$ or $\beta$ is left
unspecified. However, the choice of parameters (\ref{Poin1bAA}) is
unique in that it always corresponds to the minimal decomposition
(\ref{descri2}).
\item \label{case3}
For $\theta=\smallf{\pi}{2}$ we have
\begin{subequations}
\begin{equation}
\label{descri3}
 U = \v{R}(-\smallf{\pi}{2}) \, \v{D}(-\chi) \quad.
\end{equation}
This is a unique minimal decomposition in terms of \underline{one}
diagonal matrix (\ref{Poin1b}) and \underline{one} rotation matrix
(\ref{linPol4}).

A representation in terms of \underline{three} matrices, form
(\ref{oeq0}), may be given for parameter values
\begin{align}
 \alpha & = \beta = \smallf{\pi}{2} \quad, \quad \delta = -\chi \quad,
 \label{Poin1bBA}
\end{align}
which is unique provided that
\begin{equation}
\label{Poin1bBB}
\chi \not\in \left\{0, \smallf{\pi}{2}, \pi, \smallf{3\pi}{2} \right\}
\quad.
\end{equation}
\end{subequations}
If $\chi \in \left\{0, \smallf{\pi}{2}, \pi, \smallf{3\pi}{2}
\right\}$, a decomposition in terms of \underline{three} matrices is
not unique, since one of the angles $\alpha$ or $\beta$ is left
unspecified. However, the choice of parameters (\ref{Poin1bBA}) is
unique in that it always corresponds to the minimal decomposition
(\ref{descri3}).
\end{enumerate}
\end{theorem}
}
{\it Proof:}

We first must confirm that the right-hand sides of (\ref{oeq0},
\ref{descri2}, \ref{descri3}) indeed lie in $SU(2)$. But this is
clear, since the adjoints of these matrices are equal to their
inverses, and the determinant is equal to one in each case. So, we
know that these matrices comprise a subset of $SU(2)$; but we do not
yet know whether this subset comprises the whole of $SU(2)$. This is
what we are going to show next.

To this end we first focus on decompositions in terms of three
matrices, (\ref{oeq0}). Let $(\theta, \phi, \chi)$ be a parameter
triple pertaining to a given form (\ref{char1}). If a decomposition
(\ref{oeq0}) of $U$ exists, there must exist a triple $(\alpha, \beta,
\delta)$ such that the real and imaginary parts of each component of
the matrices (\ref{char1}) and (\ref{oeq1b}) coincide. This yields four
equations
\begin{subequations}
\label{comp1}
\begin{align}
 \cos\delta \, \cos\beta & = \ms \cos\theta \, \cos\phi \quad,
 \label{comp1a} \\
 \cos\delta \, \sin\beta & = \ms \sin\theta \, \cos\chi \quad,
 \label{comp1b} \\
 \sin\delta \, \cos\alpha & = \ms \cos\theta \, \sin\phi \quad,
 \label{comp1c} \\ 
 \sin\delta \, \sin\alpha & = - \sin\theta \, \sin\chi
 \quad. \label{comp1d}
\end{align}
\end{subequations}
We now show that the system (\ref{comp1}) is indeed solvable for the
triple $(\alpha, \beta, \delta)$, and that the solution is unique
under the circumstances described in case \ref{case1}.) --
\ref{case3}.) above.

We first examine case \ref{case1}.):

On dividing (\ref{comp1c}) by (\ref{comp1a}), and (\ref{comp1d}) by
(\ref{comp1b}), we find
\begin{equation}
\label{oeq4}
\left.
\begin{aligned}
 \tan\phi &=& &\tan\delta \; \frac{\cos\alpha}{\cos\beta} \quad, \\
 \tan\chi &=& - &\tan\delta \; \frac{\sin\alpha}{\sin\beta} \quad.
\end{aligned} \qquad
\right\}
\end{equation}
For this step to be admissible we had to use the fact that
$\cos\theta, \sin\theta \neq 0$, which is true by
assumption. Furthermore, we square and add (\ref{comp1a}) and
(\ref{comp1b}), and (\ref{comp1c}) and (\ref{comp1d}), and divide the
resulting equations. This yields the formula
\begin{equation}
\label{zw2}
 \tan^2\delta = \frac{\sin^2\phi + \sin^2\chi \,
 \tan^2\theta}{\cos^2\phi + \cos^2\chi \, \tan^2\theta} \quad,
\end{equation}
which determines $\tan\delta$ up to a sign from known quantities
$\phi$, $\chi$ and $\theta$. However, we require [see (\ref{Poin1})]
that $\sin\alpha$, $\sin\beta$ always be nonnegative. Then the second
equation in (\ref{oeq4}) implies that the sign of $\tan\delta$ is the
negative of the sign of $\tan\chi$. Thus, $\delta$ is determined up to
multiples of $\pi$. Now, since $\sin\beta$ must be nonnegative,
eq. (\ref{comp1b}) determines the sign of $\cos\delta$, and therefore
$\delta$ is uniquely determined in $[0,2\pi)$.

Now that $\delta$ is determined, $\alpha$ and $\beta$ are uniquely
determined by the system (\ref{comp1}). In fact they are determined to
lie in the interval $[0,\pi)$, for, we have used the condition
$\sin\beta \ge 0$ as well as the positivity of
$\frac{\sin\alpha}{\sin\beta}$ in (\ref{oeq4}) for the determination
of $\delta$.

Now let us examine case \ref{case2}.): Let $\theta=0$. We first treat
the case $\phi \not\in \left\{0, \smallf{\pi}{2}, \pi,
\smallf{3\pi}{2} \right\}$. Then the system (\ref{comp1}) becomes
\begin{subequations}
\label{comp2}
\begin{align}
 \cos\delta \, \cos\beta & = \cos\phi \quad, \label{comp2a} \\
 \cos\delta \, \sin\beta & = 0 \quad, \label{comp2b} \\ \sin\delta \,
 \cos\alpha & = \sin\phi \quad, \label{comp2c} \\ \sin\delta \,
 \sin\alpha & = 0 \quad. \label{comp2d}
\end{align}
\end{subequations}
Since none of $\sin\phi$, $\cos\phi$ is zero, eqs. (\ref{comp2a}) and
(\ref{comp2c}) imply that $\sin\delta, \cos\delta \neq 0$. But then
eqs. (\ref{comp2b}), (\ref{comp2d}) imply that $\sin\alpha = \sin\beta
= 0$, and due to (\ref{Poin1}) it follows that $\alpha = \beta =
0$. In this case, (\ref{comp2a}) and (\ref{comp2c}) imply that $\delta
= \phi$. This gives the minimal decomposition (\ref{descri2}), which
coincides with form (\ref{oeq0}) for the choice of parameters
(\ref{Poin1bAA}).

Now we briefly discuss the cases where $\phi \in \left\{0,
 \smallf{\pi}{2}, \pi, \smallf{3\pi}{2} \right\}$. We proceed  along
 the same lines as above, that is to say, using the system
 (\ref{comp2}) as our starting point. For $\cos\phi=0$ we arrive at
 the decomposition
\begin{subequations}
\begin{equation}
\label{cosPhi01}
 U = \v{R}(-\smallf{\beta}{2}) \v{D}(\phi) \,
 \v{R}(-\smallf{\beta}{2}) \quad,
\end{equation}
where $\alpha=0$ and $\beta$ is left unspecified. This indeterminacy
corresponds to the non-uniqueness of the parametrisation $(\theta,
\phi, \chi)$, eq. (\ref{char1}), when the parameter $\theta$ takes
values $0,\smallf{\pi}{2}$ in the boundary of the parameter
space. Eq. (\ref{cosPhi01}) contains three matrices for $\beta \neq
0$, but only one for $\beta=0$, since in this case
$\v{R}(0)=\Eins_2$. The choice $\beta=0$ therefore defines the minimal
decomposition which is unique if $\beta$ is restricted
[eq. (\ref{Poin1})] to lie in the admissible range $[0,\pi)$. This
choice of parameters coincides with (\ref{Poin1bAA}), and the
resulting form for $U$ coincides with (\ref{descri2}).

Analogously, for $\sin\phi=0$ we have the decomposition
\begin{equation}
\label{sinPhi01}
 U = \v{R}(-\smallf{\alpha}{2}) \, \v{D}(\phi) \,
 \v{R}(\smallf{\alpha}{2}) \quad,
\end{equation}
\end{subequations}
with $\beta=0$ and $\alpha$ left unspecified. For general $\alpha$,
(\ref{sinPhi01}) provides a decomposition in terms of three
matrices. A minimal decomposition, in terms of one matrix only, is
obtained for the choice $\alpha=0$, which is unique again, since
$\alpha$ is restricted to lie in $[0,\pi)$. Again, the parameter
values $\alpha = \beta = 0$ corresponding to the minimal decomposition
coincide with (\ref{Poin1bAA}), and the resulting minimal form is
(\ref{descri2}).

Now we examine case \ref{case3}.): Let $\theta = \smallf{\pi}{2}$ but
 $\chi\not\in \left\{ 0, \smallf{\pi}{2}, \pi, \smallf{3\pi}{2}
 \right\}$ at first. Then the system (\ref{comp1}) becomes
\begin{subequations}
\label{comp3}
\begin{align}
 \cos\delta \, \cos\beta & = 0 \quad, \label{comp3a} \\ \cos\delta \,
 \sin\beta & = \cos\chi \quad, \label{comp3b} \\ \sin\delta \,
 \cos\alpha & = 0 \quad, \label{comp3c} \\ \sin\delta \, \sin\alpha &
 = - \sin\chi \quad. \label{comp3d}
\end{align}
\end{subequations}
Since $\sin\chi, \cos\chi \neq 0$ by assumption, we must have
$\sin\delta, \cos\delta \neq 0$ from (\ref{comp3b}, \ref{comp3d}). But
then (\ref{comp3a}, \ref{comp3c}) imply that $\cos\alpha = \cos\beta =
0$, or $\alpha = \beta = \smallf{\pi}{2}$, due to (\ref{Poin1}). Then
(\ref{comp3b}, \ref{comp3d}) imply that $\delta = - \chi$, so all
angles are uniquely determined, and the decomposition (\ref{descri3})
follows.

Now we discuss the remaining cases, $\chi \in \left\{ 0,
 \smallf{\pi}{2}, \pi, \smallf{3\pi}{2} \right\}$. This  proceeds
 exactly as for eqs. (\ref{cosPhi01}, \ref{sinPhi01}). For
 $\cos\chi=0$ we obtain the decomposition
\begin{subequations}
\begin{equation}
\label{cosChi01}
 U = \v{R}(-\smallf{\pi/2+\beta}{2}) \, \v{D}(-\chi) \,
 \v{R}(\smallf{\pi/2-\beta}{2}) \quad,
\end{equation}
with $\alpha=\smallf{\pi}{2}$ and $\beta$ left unspecified. This
becomes minimal for $\beta=\smallf{\pi}{2}$, corresponding to
(\ref{Poin1bBA}), and leads to eq. (\ref{descri3}). For $\sin\chi=0$
we have
\begin{equation}
\label{sinChi01}
 U = \v{R}(-\smallf{\alpha+\pi/2}{2}) \, \v{D}(\chi) \,
 \v{R}(\smallf{\alpha-\pi/2}{2}) \quad,
\end{equation}
with $\beta=\smallf{\pi}{2}$ and unspecified $\alpha$. The minimal
decomposition is obtained for $\alpha=\smallf{\pi}{2}$, corresponding
to (\ref{Poin1bBA}), which leads to
\begin{equation}
\label{sinChi02}
 U = \v{R}(-\smallf{\pi}{2}) \, \v{D}(\chi) \quad.
\end{equation}
\end{subequations}
However, in the case at hand we have $\v{D}(\chi) = \v{D}(-\chi)$, so
that the minimal form is again equal to (\ref{descri3}).

This finishes the proof of theorem \ref{decoTheo}. \hfill
$\blacksquare$

\section{The \PET} \label{PoiEqu}

We are now in a position to deduce the \PET\ from our decomposition
theorem \ref{decoTheo}. Let us assume that $U$ is the Jones matrix of
a non-absorbing optical medium. As discussed before, $U$ is unitary
and may be assumed to be unimodular, possibly after discarding a
global phase. Then $U$ has a unique decomposition (\ref{oeq0}) or
(\ref{descri2}) or (\ref{descri3}). For the purposes of the \PET, we
use the decomposition (\ref{oeq0}) in terms of three matrices, so that
\begin{align}
\label{oeq111}
 U & = \v{R}(-\xi) \, \v{D}(\delta) \, \v{R}(\zeta) \quad.
\end{align}
Here we have introduced two angles
\begin{align}
\label{angles1}
 \xi  & = \frac{\alpha+\beta}{2} \quad, \quad \zeta =
 \frac{\alpha-\beta}{2} \quad,
\end{align}
and it is understood that $\xi$ and $\zeta$ are always chosen
according to the minimal decompositions (\ref{oeq0}, \ref{Poin1bAA},
\ref{Poin1bBA}). In the next step we take into account that the
diagonal matrix $\v{D}(\delta)$ is formally identical to the Jones
matrix of a linear retarder with fast axis along the $x$-direction and
a phase lag of $-2\delta$ of the slow with respect to the fast axis,
\begin{equation}
\label{oeq6}
 \v{D}(\delta) = \v{J}_{\li}(0,-2\delta) \quad,
\end{equation}
according to (\ref{ret3}). We can now rewrite (\ref{oeq111}) as
\begin{subequations}
\label{oeq7}
\begin{align}
 U & = \v{R}(-\xi) \; \v{J}_{\li}(0,-2\delta) \; \v{R}(\xi) \; 
 \v{R}(-\xi+\zeta) = \v{J}_{\li}(\xi, -2\delta) \; \v{R}(\zeta-\xi)
 \label{oeq7a} \\
 & = \v{R}(-\xi+\zeta) \; \v{R}(-\zeta) \;
 \v{J}_{\li}(0, -2\delta) \; \v{R}(\zeta) = \v{R}(\zeta-\xi) \;
 \v{J}_{\li}(\zeta, -2\delta) \label{oeq7b} \quad.
\end{align}
\end{subequations}
Here we have used (\ref{ret5}) and the group property (\ref{linPol5a})
of the rotation matrices.

In eqs. (\ref{oeq7}), the rotation matrices have been used to
transform the Jones matrices of linear retarders at different angles
of their fast axes into each other. However, in eq. (\ref{rota4}) we
pointed out that a rotation matrix $\v{R}(-\smallf{\gamma}{2})$ is
formally identical to the Jones matrix of a rotator
$\v{J}_{\ro}(-\smallf{\gamma}{2})$. Then the right-hand sides of
eqs. (\ref{oeq7a}, \ref{oeq7b}) may be rewritten as
\begin{subequations}
\label{oeq8}
\begin{align}
 U & = \v{J}_{\li}(\smallf{\alpha+\beta}{2}, -2\delta) \;
 \v{J}_{\ro}(-\beta) \label{oeq8a} \\
 & = \v{J}_{\ro}(-\beta) \; \v{J}_{\li}(\smallf{\alpha-\beta}{2},
 -2\delta) \label{oeq8b} \quad.
\end{align}
\end{subequations}

From the proof of theorem \ref{decoTheo} we know that the
decompositions (\ref{oeq7}, \ref{oeq8}) are unique provided that the
parameters $(\theta, \phi, \chi)$ in the parametrisation (\ref{char1})
of $U$ are restricted according to (\ref{spec}).

If one of the parameter sets in (\ref{spec}) is taken, the matrices in
(\ref{oeq111}) are no longer uniquely determined. From the proof of
theorem \ref{decoTheo} we know that in these cases either $\alpha$ or
$\beta$ remains unspecified. However, the phase retardation
$\Delta=-2\delta$ is always well-defined. It is easy to translate the
condition that $(\theta,\phi)$ or $(\theta,\chi)$ take one of the
values in (\ref{spec1}, \ref{spec2}), respectively, into a condition
for the optical parameters of the linear retarder and rotator in
eqs. (\ref{oeq8}). This can be done with the help of (\ref{cosPhi01},
\ref{sinPhi01}, \ref{cosChi01}, \ref{sinChi01}), or also
(\ref{descri2}) and (\ref{descri3}). It is easy to check that the
decomposition is non-unique if and only if the phase retardation
$\Delta$ of the linear retarder in the equivalent optical model is an
even or odd integer multiple of $\pi$. Physically this means that the
retarder in these cases can be realized by a stack of {\it half-wave
plates}.

In eqs. (\ref{oeq7}), (\ref{oeq8}) we have proven the
\begin{theorem}[Poincare Equivalence Theorem] \label{PoiEquTheo}
For every non-absorbing passive optical medium (represented by a
unimodular unitary matrix) there exists an optically equivalent model
which is built from one linear retarder and one rotator only. These
elements may be arranged in any order, and in both cases the phase lag
$\Delta = -2\delta$ is the same.

If $\Delta$ is not a multiple of $\pi$, $\alpha$ and $\beta$ are
uniquely determined.

If $\Delta$ is a multiple of $\pi$ (corresponding to a stack of
half-wave plates), one of the angles $\alpha$ or $\beta$ is left
unspecified. However, $\alpha$ and $\beta$ may still be chosen
according to (\ref{Poin1bAA}) and (\ref{Poin1bBA}), corresponding to
the minimal decompositions (\ref{descri2}) and (\ref{descri3}).
\end{theorem}

\section{Jones vectors, Stokes parameters, and the Poincar\'e sphere}

In eqs. (\ref{Jones2}, \ref{Jones3}) we have defined the polarization
form and the associated Jones vector of a completely polarized beam of
light. We mentioned that these quantities are defined only up to a
global phase, so that Jones vectors $v$ and $e^{i\Phi}\, v$ are
equivalent.

An alternative description of the light beam in terms of the so-called
{\it Stokes parameters} removes this ambiguity: The reason is that, in
contrast to the Jones vector, the Stokes parameters are quadratic in
the (complex) field strength and can, in principle, be determined by
four intensity measurements, in conjunction with a linear polarizer
and a quarter-wave plate, or equivalents \cite{BornWolf, Jackson}. The
Stokes parameters can be motivated by observing that, for a wave
propagating in the $z$-direction, the scalar products $\v{e}_1 \bl
\v{E} = a_1$ and $\v{e}_2 \bl \v{E} = a_2$ are the complex amplitudes
of electric field strength linearly polarized in the $x$- and
$y$-direction, respectively. The phase difference between these
components accounts for the (in general) elliptical polarization form
of the light beam. Then the Stokes parameters are defined in terms of
these quantities as
\begin{equation}
\label{Stokes1}
\left.
\begin{aligned}
 s_0 & = \left| \v{e}_1\bl\v{E} \right|^2 + \left| \v{e}_2\bl\v{E}
 \right|^2 = a_1^2 + a_2^2 \quad,\\
 s_1 & = \left| \v{e}_1\bl\v{E} \right|^2 - \left| \v{e}_2\bl\v{E}
 \right|^2 = a_1^2 - a_2^2 \quad,\\
 s_2 & = 2\, \Re \left[ \left(\v{e}_1\bl\v{E}\right)^*
 \left(\v{e}_1\bl\v{E}\right) \right] = 2\, a_1\, a_2\, \cos(\delta_2 -
 \delta_1) \quad,\\
 s_3 & = 2\, \Im \left[ \left(\v{e}_1\bl\v{E}\right)^*
 \left(\v{e}_1\bl\v{E}\right) \right] = 2\, a_1\, a_2\, \sin(\delta_2 -
 \delta_1) \quad.
\end{aligned} \qquad
\right\}
\end{equation}
The electric field vector of the polarized light beam, expressed by
(\ref{Jones3}) or (\ref{Stokes1}), describes an ellipse whose
principal axes are not in the $x$- and $y$-direction in general. The
first principal axis, with length $2a$, makes an angle $\psi$ with the
$x$-axis, while the second principal axis, being perpendicular to the
first, has a length $2b$. The shape and orientation of the ellipse can
be described by the angle $\chi$ such that $\tan \chi = \pm b/a$. It
is then a matter of elementary computation to show that the Stokes
parameters (\ref{Stokes1}) can be expressed in terms of these angles
as \cite{BornWolf}
\begin{equation}
\label{Stokes2}
\left.
\begin{aligned}
 s_1 & = s_0\, \cos 2\chi\, \cos 2\psi \quad, \\
 s_2 & = s_0\, \cos 2\chi\, \sin 2\psi \quad, \\
 s_3 & = s_0\, \sin 2\chi \quad.
\end{aligned} \qquad
\right\}
\end{equation}
Evidently, the system (\ref{Stokes2}) can be interpreted as the
Cartesian coordinates of a point on a sphere of radius $s_0$, the
so-called {\it Poincar\'e sphere} \cite{Poincare1892}. The
significance of the Poincare sphere lies in the fact that its points
are in 1--1 correspondence with the distinct states of polarization of
a completely polarized light beam with constant intensity; the
ambiguity in the Jones matrix formalism, in which a continuous
infinity $e^{i\Phi}\, v$ of Jones vectors corresponded to physically
equivalent polarization forms, has been removed. One can show that
right- (left-) handed circular polarization is represented by the
north (south) pole on the Poincar\'e sphere; whilst the states of
linear polarization correspond to the points in the equatorial
plane. All other points describe a general elliptical
polarization. Each of the equatorial points represents a different
angle of the plane of linear polarization; as a consequence, an
optical rotator, changing just this angle, but preserving the linear
polarization form, may be regarded as a rotation $R_3$ about the
$z$-axis on the Poincar\'e sphere. Similarly, an optical retarder,
capable of transforming a linear polarization state into a general
elliptical one, will correspond to a rotation about an angle $2\psi$
whose axis lies in the $xy$-plane of the coordinate system. However,
any such rotation can be obtained as a sequence $R_3\, R_2\, R_3^{-1}$
of rotations about the $z$-axis, the momentary $y$-axis, and the new
$z$-axis again.

More generally, it can be shown that the action of any loss-less
optical element corresponds to a proper rotation on the Poincar\'e
sphere. From this fact, the \PET\ can be inferred immediately: For,
let $U$ denote the unitary matrix of the optical element in the Jones
matrix formalism, and let $R(U)$ denote the corresponding rotation on
the Poincar\'e sphere. We know that every proper rotation can be
decomposed into three {\it Euler rotations}, which may be chosen as a
sequence $R'_3\, R_2\, R_3$ of three rotations about the momentary
$z$- and $y$-axes. From the previous paragraph we know that the
factors in this decomposition represent the actions of rotators and
linear retarders, so that the sequence $R'_3\; R_2\, R_3$ corresponds,
in the Jones matrix formalism, to a sequence of optical elements
$\v{R}'\, \v{D}\, \v{R}$, which is of course just eq. (\ref{oeq0}). In
this way, the decomposition of $SU(2)$-matrices into 'basic building
blocks' as in the \PET\ can be traced back to the Euler decomposition
of proper rotations on the Poincar\'e sphere.

The parameters in (\ref{Stokes1}) are classical quantities, of
course. It is possible to define associated quantum Stokes parameters
together with a quantum Poincar\'e sphere; these concepts are reviewed
in \cite{Leonhardt-ROP66}, for example. In \cite{KorolkovaEA-PRA65},
an application of the quantum Stokes parameters to squeezed light has
been presented.

\section{Relation between Characteristic Parameters and the
Poincar\'e-Equivalent Model}
\label{Relation}

In section \ref{CharacteristicParameters} we have shown that every
optical device which (in the framework of Jones calculus) can be
represented by a unitary unimodular $(2,2)$ matrix $U$ possesses two
orthogonal directions, $w_1$ and $w_2$, such that linearly polarized
light whose plane of polarization on entry into the apparatus is $w_1$
or $w_2$ will emerge linearly polarized again, with polarization
directions $w'_1$ and $w'_2$. Furthermore, we saw that linearly
polarized light entering at the two primary characteristic directions
emerges with different phases; an expression for the relative phase
difference was given in (\ref{Phdiff0}), but we were not able to
determine it uniquely.

On the other hand, in theorem \ref{PoiEquTheo} we showed that every
such device was optically equivalent to a succession of linear
retarders and rotators. Since a linear retarder has two preferred
spatial directions (the fast and slow axis) and a well-defined phase
retardation between them, this suggests that the characteristic
directions might be closely related to the fast and slow axes, while
the phase difference (\ref{Phdiff0}) might be identified with the
retardation of the retarder.

We now show that this is indeed so: We first treat the unique cases,
i.e. parameters $(\theta, \phi, \chi)$ satisfy (\ref{spec}), or
equivalently, the phase lag $\Delta=-2\delta$ of the equivalent linear
retarder is not a multiple of $\pi$. In this case the linear retarder
has precisely two preferred directions, namely the fast and slow
optical axis. Only at these directions preserves the linear retarder
the state of linear polarization; any other direction of linear
polarization on entry will emerge as elliptical polarization form. On
using the decomposition (\ref{oeq111}) it is then clear that the
primary characteristic directions are determined by the condition
\begin{equation}
\label{cond1}
\left.
\begin{aligned}
 \v{R}(\zeta) \, w_m = e_m \quad & , \quad m=1,2 \quad, \\
  e_1 = (1\, ,\,0)^T \quad & , \quad e_2 = (0\,,\,1)^T \quad.
\end{aligned} \qquad
\right\}
\end{equation}
Therefore we must have
\begin{equation}
\label{cond2}
 w_1 = \v{R}(-\zeta) \, e_1 = \left[ \begin{mat}{1}
 &\cos\left(\smallf{\alpha-\beta}{2} \right) \\
 &\sin\left(\smallf{\alpha-\beta}{2} \right)\end{mat} \right] 
 \quad, \quad
 w_2 = \v{R}(-\zeta) \, e_2 = \left[ \begin{mat}{1}
 -&\sin\left(\smallf{\alpha-\beta}{2} \right) \\
 &\cos\left(\smallf{\alpha-\beta}{2} \right)\end{mat} \right] \quad,
\end{equation}
where we have {\it defined} that $w_1$ be the direction corresponding
to the {\it faster} phase velocity. Similarly, the secondary
characteristic directions must be
\begin{equation}
\label{cond3}
 w'_1 = \v{R}(-\xi) \, e_1 = \left[ \begin{mat}{1}
 &\cos\left(\smallf{\alpha+\beta}{2} \right) \\
 &\sin\left(\smallf{\alpha+\beta}{2} \right)\end{mat} \right]
 \quad, \quad
  w'_2 = \v{R}(-\zeta) \, e_2 = \left[ \begin{mat}{1}
 -&\sin\left(\smallf{\alpha+\beta}{2} \right) \\
 &\cos\left(\smallf{\alpha+\beta}{2} \right)\end{mat} \right] 
 \quad.
\end{equation}
Again, we have {\it defined} that $w'_1$ be the secondary
characteristic direction corresponding to the {\it faster} phase
velocity. A comparison between (\ref{cond2}) and (\ref{char2}), and
(\ref{cond3}) and (\ref{char3}) now shows that we should set
\begin{subequations}
\label{angles2}
\begin{equation}
\label{angles2a}
 \gamma_1 = \smallf{\alpha-\beta}{2} = \zeta \quad, \quad \gamma_2 =
 \gamma_1 + \smallf{\pi}{2} \quad, 
\end{equation}
and
\begin{equation}
\label{angles2b}
 \gamma'_1 = \smallf{\alpha+\beta}{2} = \xi \quad, \quad \gamma'_2 =
 \gamma'_1 + \smallf{\pi}{2} \quad,
\end{equation} 
\end{subequations}
where we have used (\ref{angles1}). 

Light which travels through the medium with linear polarization along
$w_1$, $w_2$ acquires a phase (relative to a global phase $\Phi_g$) of
\begin{equation}
\label{cond4}
 \Phi_1 = \delta \quad, \quad \Phi_2 = -\delta \quad.
\end{equation}
Then the total phase difference between directions $w_1$ and $w_2$ is
\begin{equation}
\label{cond5}
\Delta = \Phi_2 - \Phi_1 = -2 \delta \quad.
\end{equation}

The arguments leading to eqs. (\ref{cond2}--\ref{cond5}) clearly are
based on intuition rather than rigorous derivation. In order to be
rigorous we have to show that these equations are consistent with the
results of theorems \ref{CharacteristicDirections},
\ref{DeterminationPhases} and eq. (\ref{Phdiff0}) in section
\ref{CharacteristicParameters}. More precisely, we must show that the
angles $\gamma_m$ as defined in (\ref{angles2a}) satisfy the equations
(\ref{char4}), while $\gamma'_m$ as defined in (\ref{angles2b}) must
satisfy (\ref{char6}). This can be done easily, on using the
assumption that parameters $(\theta, \phi, \chi)$ satisfy
(\ref{spec}): To show that the first statement is true it is
sufficient to expand $\tan(\alpha-\beta)$ in terms of sines, cosines
of $\alpha$ and $\beta$, and multiply both nominator and denumerator
of the resulting fraction with $\cos\delta\sin\delta$; it is at this
point where we need the condition that $\cos\delta, \sin\delta \neq
0$. The result can be easily recast in the form of
eq. (\ref{char4}). The second statement linking (\ref{angles2b}) to
(\ref{char6}) is confirmed in the same way. Finally,
eqs. (\ref{char3ba}, \ref{char3bb}) can be derived immediately from
the system (\ref{comp1}). We note that we have removed the
indeterminacy concerning the assignment between $\Phi_m$ and
$\pm\delta$ by decreeing that the subscript $1$ should refer to the
fast axis of the equivalent retarder. This assignment is now
well-justified since it is based on the distinction between different
values of a physically measurable quantity, namely the phase velocity
along the axes. Previously, in section \ref{CharacteristicParameters},
we did not yet possess the physical justification for such a
deliberate choice.

We now turn to discuss the non-unique decompositions: From section
\ref{PoiEqu} we know that in these cases the phase retardation of the
equivalent linear retarder is always a multiple of $\pi$; in
particular, it is always well-defined. What is left unspecified is one
of the angles $\alpha, \beta$. This is physically reasonable, for,
whenever $\Delta=k\pi$, $k \in \ZZ$, the retarder transforms {\it any}
linearly polarized light into linearly polarized light again, and
hence any direction (before entry into the medium) is a primary
characteristic direction (and any direction behind the medium is a
secondary one). This is clearly consistent with the eqs. (\ref{char4},
\ref{char6}): In the comment following (\ref{spec}) we remarked that
(\ref{char4}, \ref{char6}) remained ill-defined if parameter values
(\ref{spec}) were taken. This simply means that there are no preferred
directions in this case, and any two orthogonal directions before and
behind the medium may serve as primary and secondary characteristic
directions.

We finish this section by writing down the decompositions (\ref{oeq8})
of the equivalent optical model in terms of the characteristic
parameters: In the case of unique decompositions, eqs. (\ref{oeq111})
and (\ref{oeq8}) read
\begin{subequations}
\label{char200}
\begin{align}
 U & = \v{J}_{\ro}(-\gamma'_1) \; \v{J}_{\li}(0,\Delta) \;
 \v{J}_{\ro}(\gamma_1) \label{char200a} \\
 & = \v{J}_{\li}(\gamma'_1,\Delta) \; \v{J}_{\ro}(\gamma_1-\gamma'_1)
 \label{char200b} \\ 
 & = \v{J}_{\ro}(\gamma_1-\gamma'_1) \; \v{J}_{\li}(\gamma_1,\Delta)
 \quad, \label{char200c}
\end{align}
where we have used (\ref{angles2}) and (\ref{cond5}). For $\theta=0$,
the minimal decomposition (\ref{descri2}) can be expressed as
\begin{equation}
\label{char200d}
 U = \v{J}_{\li}(0,\Delta) \quad, \quad \Delta = k \pi \quad, \quad
 k\in\ZZ \quad.
\end{equation}
For $\theta=\smallf{\pi}{2}$, the minimal decomposition
(\ref{descri3}) can be expressed as
\begin{equation}
\label{char200e}
 U = \v{J}_{\ro}(-\smallf{\pi}{2}) \; \v{J}_{\li}(\Delta) \quad, \quad
 \Delta = k\pi \quad, \quad k\in \ZZ \quad.
\end{equation}
\end{subequations}

\section{Basic Ideas of Integrated Photoelasticity}
\label{Photoelasticity}

In this section we want to outline, in a single paragraph, how optical
media with spatially dependent dielectric tensors naturally emerge in
Photoelasticity.

Consider a medium (such as glass or certain polymers) which is
optically homogeneous and isotropic, with a dielectric tensor
$\epsilon\delta_{ij}$ when strain-free, but which becomes
inhomogeneous and anisotropic when subject to an external load. The
dielectric tensor $\epsilon_{ij}$ then is related to the stress tensor
$\sigma_{ij}$ of the material according to the {\it stress-optical
law}
\begin{equation}
\label{stress1}
 \epsilon_{ij} = \epsilon \, \delta_{ij} + C_1 \, \sigma_{ij} + C_2 \,
 \tr(\sigma) \, \delta_{ij} \quad, \quad i,j=1,2,3 \quad,
\end{equation}
where $C_1, C_2$ are stress-optical constants. The medium is assumed
to occupy a bounded region in space. Polarized light is sent through
the specimen at many different angles, and from the change of the
state of polarization the characteristic parameters of the model, for
a given direction of the incident light beam, are retrieved. Light is
assumed to propagate through the medium along straight lines, an
approximation which is acceptable as long as the stress-induced
birefringence is weak. Under these circumstances, the Maxwell
equations describing the propagation of a plane wave with locally
varying state of polarization along the $z$-direction of the medium
can be cast into the form
\begin{equation}
\label{bas10}
 \frac{d}{d z} \left( \begin{mat}{1} &v_1\\ &v_2 \end{mat} \right) =
 \frac{i \omega}{2c\sqrt{\epsilon_0 \epsilon}} \left( \begin{mat}{2}
 &\epsilon_{11}-\epsilon & & \epsilon_{12} \\ &\epsilon_{21} & &
 \epsilon_{22}-\epsilon \end{mat} \right) \left( \begin{mat}{1} &v_1\\
 &v_2 \end{mat} \right) \quad,
\end{equation}
where $v_1(z), v_2(z)$ are the components of the spatially dependent
Jones vector for fixed $x$ and $y$, and $\epsilon_{ij}(z)$, $i,j={\bf
1,2}$ are the components of the projection of the local dielectric
tensor onto the $xy$-plane, at position $(x,y,z)$. The point $(x,y)$
in the $xy$-plane characterizes the location of entry of the light
beam (which is assumed to have very small diameter) into the
medium. Eqs. (\ref{bas10}) may be expressed in an equivalent way by
introducing a unitary propagation matrix $U(z)$ such that
\begin{equation}
\label{bas11}
 v(z) = U(z,z_i) \, v(z_i) \quad,
\end{equation}
where $z_i$ is any point before the medium. This matrix must
satisfy a differential equation analogous to (\ref{bas10}),
\begin{equation}
\label{bas12}
 \frac{d}{d z} \, U(z,z_i) = i\, H(z) \; U(z,z_i) \quad,
\end{equation}
where the matrix $H$ is given on the right-hand side of
(\ref{bas10}). If $z_f$ is a point behind the medium, the matrix
$U(z_f,z_i)$ is obtained by integrating (\ref{bas12}); it determines
the change of polarization form of the light beam passing through the
medium between points $(x,y,z_i)$ and $(x,y,z_f)$ along a straight
line. Measurements of this change in the polarization determine the
characteristic parameters, and in turn, the equivalent model in either
of the forms (\ref{char200}), for the given ray of light. Repeating
the procedure for many rays at different angles yields data sets which
contain the characteristic parameters for any possible direction of
the light beam. From these data we attempt to reconstruct the local
dielectric tensor in the interior of the medium. This problem, called
{\it three-dimensional Photoelasticity}, is still unsolved in its most
general form, and the \PET\ as given in one of the forms above may
play an important role in this goal.

\section{Summary}
\label{Summary}
In this paper we have proven a decomposition theorem for
two-dimensional unitary unimodular matrices which represent a
non-absorbing passive optical medium in the framework of the Jones
calculus. From this decomposition theorem, the \PET\ can be derived as
a corollary. The latter states that every non-absorbing optical medium
is optically equivalent to a succession of linear retarders and 
rotators, in any order. Except for few cases, the equivalent linear
retarder and rotator are uniquely determined from data characterizing
the original medium. We have shown that the parameters of these
equivalent elements have a simple interpretation in terms of
measurable quantities, the so-called Characteristic Parameters of an
optical model, which have been known in the engineering literature for
some time. The relation between these Characteristic Parameters and
the Poincar\'e-equivalent optical model has been fully
clarified. Finally, we have illustrated the context in which the
Poincar\'e-equivalent model is expected to become a useful tool:
Within the framework of three-dimensional Photoelasticity, the
decomposition theorems proven in this paper may turn out to be
important to the goal of reconstructing local dielectric tensors from
measurements of the global change of the polarization state of light
transmitted through an inhomogeneous and anisotropic medium with
spatially varying refractive tensor.

\section*{Acknowledgements}
 Hanno Hammer wishes to acknowledge acknowledge support from EPSRC
 grant GR/86300/01.


\end{document}